# Features Matching Using Natural Language Processing

Muhammad Danial Khilji

Department of Data Science, choreograph

## Abstract

*The feature matching is a basic step in matching different datasets. This article proposes shows a new hybrid model of a pretrained Natural Language Processing (NLP) based model called BERT used in parallel with a statistical model based on Jaccard similarity to measure the similarity between list of features from two different datasets. This reduces the time required to search for correlations or manually match each feature from one dataset to another.*

## 1. Introduction

The features matching is a first step in many of the data processes. It plays a crucial part in data fusion process. Data fusion is the process of populating one dataset with another one. Schema matching if performed manually, is considerably time consuming especially when the width of data is huge. In this article, language-based approach is built for feature matching. It is mostly used when the datasets being matched are heterogenous in nature.

The section 2 of this paper presents previous work published for schema matching and then it highlights how machine learning advancement changed the shape of schema matching. There is a thorough review of BERT related models that perform well in similarity matching and can be potentially used for schema matching as well. The section 3 explains the architecture in-depth that is designed and implemented in Python language. It starts by bringing the data into the right structure and applying some text cleaning to make it easier to understand for similarity measures in section 3.1-3.3. The section 3.4 elaborates on the two different methods that have been used to measure the similarity between two feature names. Then section 3.5 gives a description of steps taken to bring the results into form easier to understand and access. A discussion on the results, in section 4, which also shows reasons on how results can vary based on type of data used. The section 5 summarizes methodology and discusses future improvements.

## 2. Literature Review

The research on feature matching is still considerably narrow considering the range of applications in which feature matching is required. The paper on Schema Matching with Opaque Column Names is one of the earliest research on features matching between two different datasets. This explains a way to matching column names and column values as well. It focuses on un-interpreted structure matching method in which column values are matched element wise in contrast to structure matching [1]. Prior to this paper, a thorough research was done by E. Rahm that shows the landscape of approaches for schema matching [2]. The major constraint in earlier approaches is that they designed strictly for databases [3]. The current popular model for schema matching is COMA which uses graph network approach to find similar features between multiple datasets [4]. There is a recent study of schema matching approaches with graph network approaches that compares the performance of COMA with other advanced ML models [5].

There has been research on schema matching approaches [6]. Most of the approaches suffer from understanding the semantic similarity between two sentences. This paper gives a brief mention of Schema-Based, Instance-Based and ML-Based approaches separately. Although, ML-Based approaches outperform other types of approaches generally, but it also depends on specific use cases. It also shows a new end-to-end schema matching approach using combination of above-mentioned types of approaches and compares it with Google's Universal Serial Encoder (USE) model [7].

With the recent advancements in machine learning field, the Google AI Language team developed the Bidirectional Encoder Representation from Transformers (BERT) model which produced state-of-the-art results in wide variety of NLP tasks [8]. Since then, the BERT model has been used to improve the entity matching in many different aspects [9]. The nature of feature matching is such that it can be classified as a branch of entity matching.

There are some useful research papers for entity matching which can also be used for feature matching. "Entity Matching with Transformer architectures" is the earliest paper that uses BERT [9], XLNet [10], RoBERTa [11] and DistilBERT [12] and compare them with classical deep-learning models and proves that transformer architecture outperforms them all. Another earlier use of BERT model in similarity matching is used for product matching from different retailers [13]. It uses more narrowed variations of state-of-the- art BERT model, BERT [8] and DistilBERT [12], for matching products using datasets from different e-commerce websites. It fine tunes the standard BERT model with an additional layer and named it eComBERT. It then compares all models where it shows that eComBERT gives considerably improved results on different datasets.

There was a successful effort on developing an architecture specifically for entity resolution named BERT-ER [14]. It delays the pair-wise interaction through enhanced alignment network and incorporating a 'blocking decoder' module that removes any obvious dissimilar matches. The results show improved F1 score at shorter speed.

Right after this, an- other paper was published which shows a different approach which uses binary matching and multi-class classification [9]. This way, the final decision during training is result of match/non-match decision and prediction of the entity identifier for each entity description hence called Joint BERT. There exist approaches other than using Neural Network models such as SERIMI which suggests class-based matching technique [15, 16].

## 3. METHODOLOGY

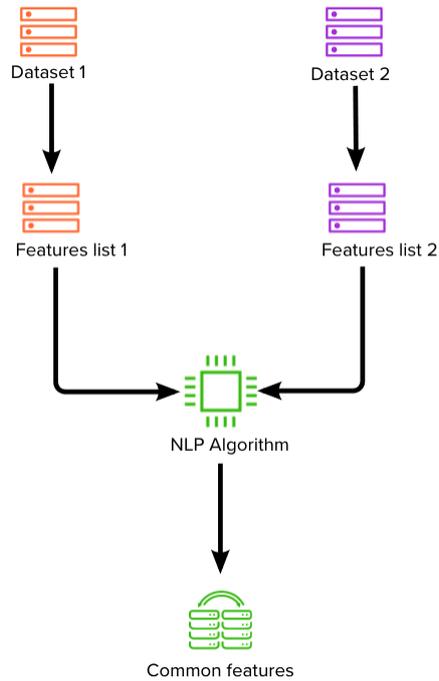

Figure 1: Main workflow

### 3.1. Pre-Processing

Firstly, the features from two different datasets are saved in two separate lists. For now, any datasets can be assumed, the results with example datasets are discussed in section 4. Both lists of features are concatenated with extra information that explains the feature to give more context which eventually improves similarity results. This extra information is explanation of each feature. Alternatively, unique feature values can be concatenated with the feature name. To remove unwanted data, these two feature lists are then pre-processed to clean the data removing null records and columns that may not be required

### 3.2. Natural Language Processing

Natural language processing is done to bring text into similar structure. The first step is to convert the text into lower case. Then remove special characters that may cause low quality results such as & % $ etc. Every sentence is then tokenized and the resulting list of tokens for each sentence is saved in separate position of another list. This way, final list will have lists of tokenized sentences. The stop words are also removed at this stage such as 'is, a, are, as'. The last step lemmatization which converts the word forms into present-tense form such as "pointing" to "point", "swimming" to "swim" [17].

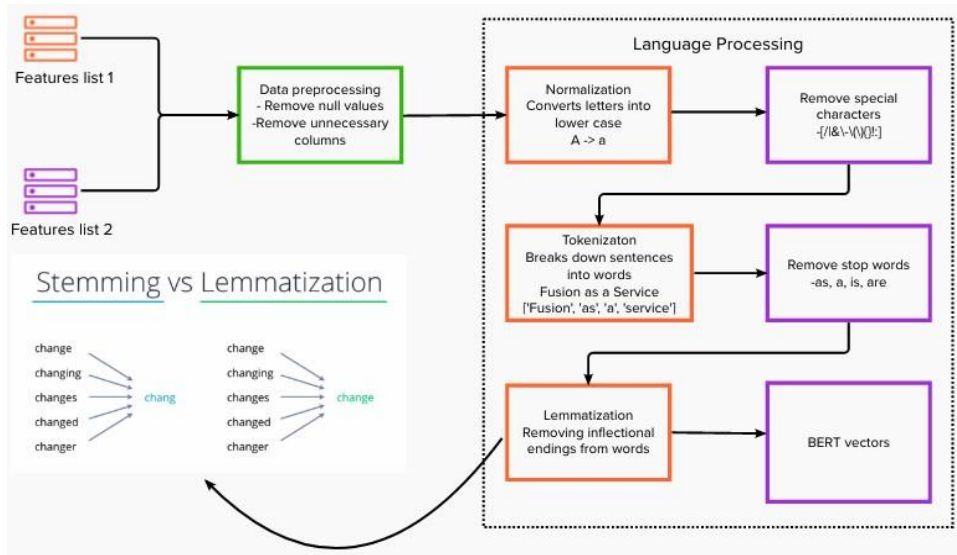

Figure 2: Data input and pre-processing stage

## 3.3. Embeddings

Once the words are lemmatized, their embeddings are generated using the pretrained standard BERT model [8]. JointBERT was also used initially but it performed poorly therefore rejected [9].

## 3.4. Similarity Measures

There are two approaches that were used in parallel, Cosine and Jaccard using BERT embeddings and using lemmatized tokens as an input respectively. Main purpose of using two similarity methods was that if one match is given low importance by one method then the other method might give it a higher importance giving conflicting similarity results. Although BERT model can perform well on its own, but Jaccard gives more flexibility to the overall model and performs for cases where special symbols or jargon is used. The Jaccard similarity working is such that for each feature match, Jaccard similarity takes the set of lemmatized tokens of records from two different lists of features and calculates the common words between them [18]. For example:

Feature from list 1: AI is our friend, and it has been our friend
Feature from list 2: AI and human has always been our friends

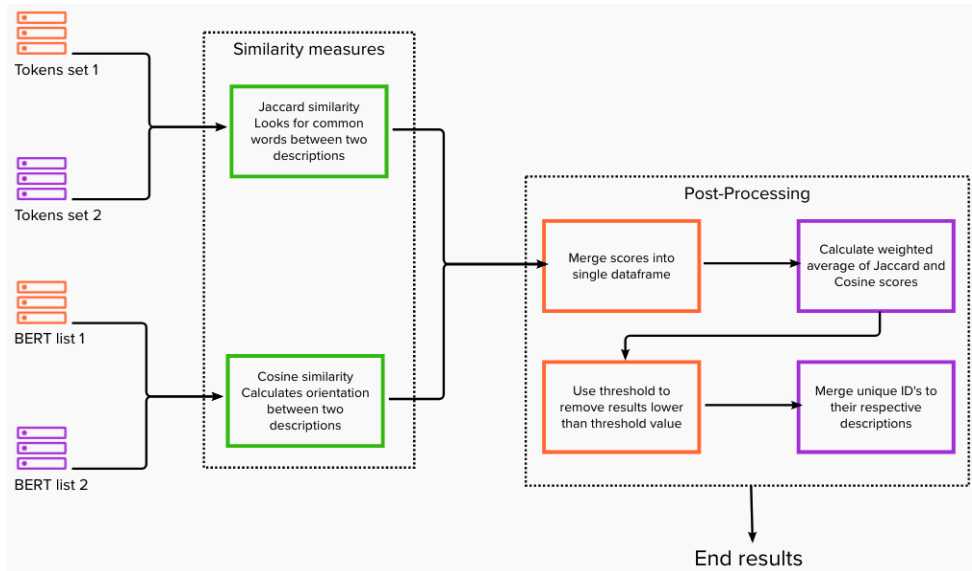

Figure 3: Pre-processing stage output to similarity measures and post-processing

$$J(A, B) = \frac{|A \cap B|}{A \cup B}$$

Where $|A \cap B|$ is the sum of common tokens, and $|A \cup B|$ is the sum of all tokens from both feature lists. A is set of tokens of individual feature from first feature list and B is set of tokens of individual feature from second feature list. The result is shown in the figure 4 below.

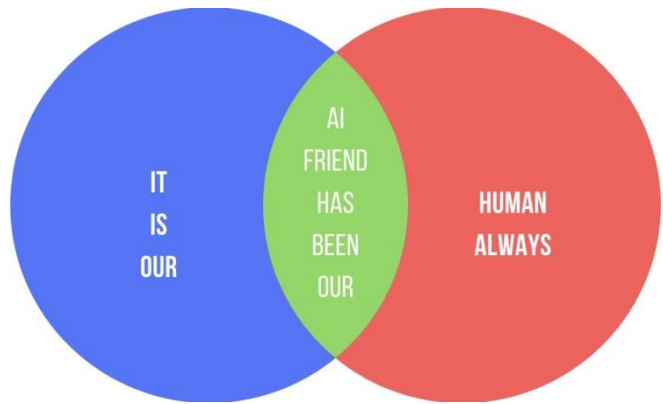

Figure 4: Venn diagram showing common words identified by Jaccard similarity (in green region)

BERT is a neural network architecture specifically designed and trained for natural language processing related tasks. It understands the similar words and gives similar numerical embedding for words which have very similar meaning [8]. The major improvement in this model is due to the model's ability to understand text bidirectionally in contrast to previous models that only understand text from one direction. The Cosine similarity takes the BERT vectors as an input from both lists of features [19]. It then plots all these features in an n-dimensional space such that closely related vectors angled close to each other. It calculates the orientation between each two vectors and gives this similarity as a result as shown in the figure 5 below using the formula:

$$Similarity = cos\theta = \frac{A.B}{||A||||B||}$$

Where A and B are set of BERT embeddings for individual features from first and second feature list respectively.

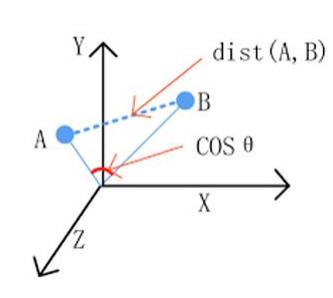

Figure 5: Cosine similarity. Note dist(A,B) is ||A||||B|| in the equation above [20]

## 3.5. Post Processing

Once the similarities are measured, they are converted into data frames, and which are then merged using original feature names. The weighted average of Jaccard and Cosine scores is calculated. The weights are assigned manually but can be improved by using logic in the future. In the algorithm, three choices are given:

$$Cosine = 1 - Jaccard$$

**Option 1:** Cosine = 0.7 and Jaccard = 0.3 (default)
**Option 2:** Cosine = Jaccard = 0.5
**Option 3:** Cosine = 0.3 and Jaccard = 0.7

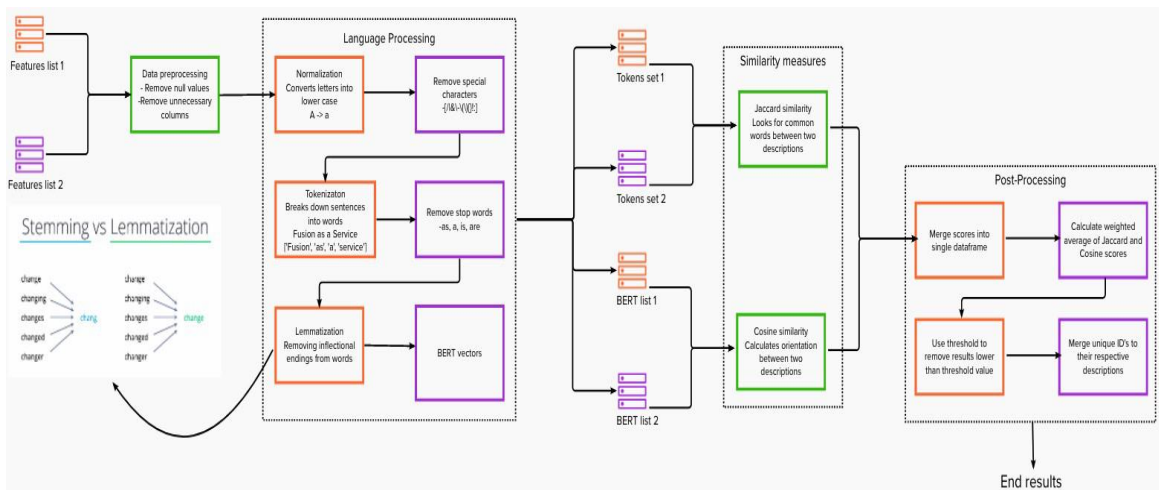

Figure 6: End-to-end workflow in detail

There may be many matches with very low score or no match at all, therefore a default some threshold value is used. It is noticed that threshold of 0.7 works the best, however user is given freedom to input threshold value of their choice within 0-1 range. Scores above 0.7 are mostly the closest matches therefore it is used as a default value although user can use lower threshold value to see weak matches. This way, all the matches that have score less than threshold value will be deleted. The last step is optional, if there are unique ID's for features then they can also be merged at this stage.

For both methods, all the features from one list are compared with all the features from the other list as shown in the figure 7 below.

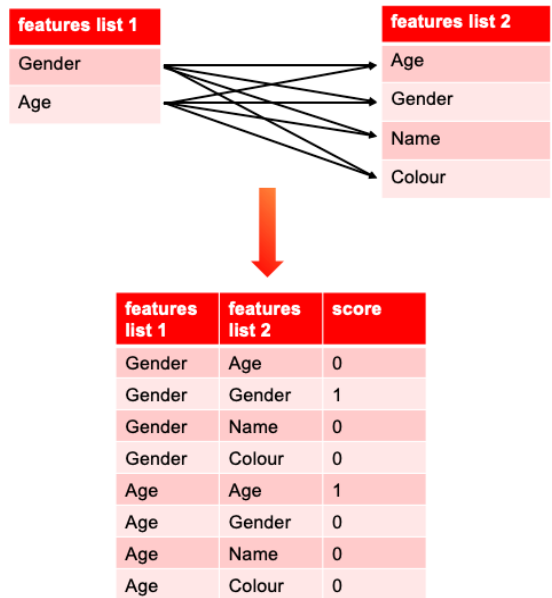

Figure 7: Features matches example

## 4. RESULTS

Most of the results were generated for data that cannot be released to public therefore, for demonstration of this paper, public datasets are used. Two test results are generated that match between IMDB and Netflix data and between Perth and King County houses sale data downloaded from Kaggle website [2, 21–23].

Table 1: Features match between Perth and King County house sales data (threshold = 0.7) [2,23]

| Perth Metropolitan Australia Features | King County USA Features | Weighted Average Score |
|---|---|---|
| PRICE | price | 1 |
| BEDROOMS | bedrooms | 1 |
| BEDROOMS | lat | 0.699999988 |
| PRICE | lōng | 0.699999988 |
| SUBURB | sqft living15 | 0.699999928 |
| LANDAREA | yrbuilt | 0.643930674 |
| SUBURB | condition | 0.595897615 |

Table 2: Features match between IMDB and Netflix data (threshold = 0.7) [21, 22]

| IMDB Features | Netflix Features | Weighted Average Score |
|---|---|---|
| Director | director | 0.99999999 |
| Released Year | Release year | 0.73863158 |
| Series Title | title | 0.712217 |
| IMDB Rating | rating | 0.65647271 |
| Overview | description | 0.65626079 |
| Overview | Listed in | 0.63430017 |
| Overview | type | 0.62954128 |

In table 1, the top two matches give a score of 1 as they are both understood as same by the BERT and both are common words with same spellings therefore Jaccard similarity also gives 1 as a result. The rest of the matches are not the best matches because both datasets do not contain same kind of features. In this scenario, BERT tries to match words that may have similar meaning such as 'Suburb' is 'Condition'. Although, both words have entirely different meaning, but BERT understands them as similar.

The table 2 shows a relatively different behavior. Although, top two matches are perfect matches, but the BERT understands them as different which gives an average lower score. In contrast, Jaccard similarity gives a score of 1 for the two matches due to which the final weighted average score is increased. Such cases are the main reason to use a simple similarity method against complex neural network models. JointBERT was also used initially but it showed poor performance on our use case. Training our own model can be considered but JointBERT requires relatively more data to learn compared to standard BERT model [17].

A pre-trained model is used for this research, but model can be trained on specific data. Although, such models have a higher chance of facing bias among other issues which can be improved using Selection Bias Exploration and Debias methods [24].

## 5. CONCLUSION

This article shows application of BERT against a simple statistical method to find similar meaning feature names. This process is useful in data fusion among other things. More than two similarity measures can be used in parallel such as Monge- Elkan and Soft TF-IDF etc [25–27]. Although more methods will give better results, but the algorithm will get computationally expensive. Therefore, only two methods have been used that suit well to our needs. The general recommendation will be to use Cosine similarity with any other methods that gives best results. Major factors that influence the results are the type of similarity that best suits the problem such as this problem falls under Token Based and Sequence Based similarity groups [28].

ACKNOWLEDGEMENTS

I would like to thank the Data Science department of Choreograph for giving me an opportunity to research and develop this methodology. My manager at the time, Victor Moron Tejero ensured I have all the resources to carry the research. Fusion-as-a-Service (FaaS) team gave me a better exposure to the problem and helped me improve the design by constant feedback. Giuseppe De Vitis, who gave me different perspectives and convinced to further improve the architecture. Nicholas Iles, as my current manager helped me get to the right people whenever needed. Georgios Giasemidis, who helped me a lot to get this research to be published by guiding me to the right journals and teaching me to structure the research paper. I would like to thank each one of you wholeheartedly.

International Journal on Cybernetics & Informatics (IJCI) Vol. 12, No.2, April 2023

**AUTHOR**

**Muhammad Danial Khilji** is a data scientist by profession with just under a year of experience in the field. He has done Masters in Applied Data Sciences and Bachelors in Electrical Engineering with majors in Artificial Intelligence and Machine Learning.

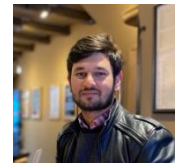